\documentclass[aps,prb,twocolumn,floatfix,showpacs]{revtex4-1}
\usepackage{graphicx}
\usepackage{amsmath,bm,amssymb,amsfonts}

\begin{document}

\title{Helical currents in metallic Rashba strips.}

\author{Ignacio J. Hamad, Claudio J. Gazza, and Jos\'e A. Riera}
\affiliation{Instituto de F\'{\i}sica Rosario, Consejo Nacional de 
Investigaciones Cient\'ificas y T\'ecnicas, and
Universidad Nacional de Rosario, Rosario, Argentina
}

\date{\today}

\begin{abstract}
We study the texture of helical currents in metallic planar strips in
the presence of Rashba spin-orbit coupling (RSOC) on the lattice at
zero temperature. In
the noninteracting case, and in the absence of external electromagnetic
sources, we determine by exact numerical diagonalization of the 
single-particle Hamiltonian, the distribution across the strip section of
these Rashba helical currents (RHC) as well as their sign oscillation,
as a function of the RSOC strength, strip width, electron filling,
and strip boundary conditions. Then, we study the effects of charge
currents introduced into the system by an Aharonov-Bohm flux for the
case of rings or by a voltage bias in the case of open strips. The 
former setup is studied by variational Monte Carlo, and the later, by
the time-dependent density-matrix-renormalization group technique.
Particularly for strips formed by two, three and four coupled chains,
we show how these RHC vary in the presence of such induced charge
current, and how their differences between spin-up and spin-down
electron currents on each chain, help to explain the distribution
across the strip of charge currents, both of the spin conserving and
the spin flipping types. We also predict the appearance of 
polarized charge currents on each chain. Finally, we show that these
Rashba helical currents and their derived features remain in the
presence of an on-site Hubbard repulsion as long as the system 
remains metallic, at quarter filling, and even at half-filling 
where a Mott-Hubbard metal-insulator transition occurs for large
Hubbard repulsion.
\end{abstract}


\maketitle

\section{Introduction}
\label{introsection}

There is an ongoing effort to take advantage of electron spin, in 
addition to its charge, for computing and information technologies,
that has evolved into the new field of 
spintronics.\cite{prinz,wolf,zutic} An important ingredient for spintronics
devices, is the spin-orbit (SO) interaction, which provides the
essential mechanisms for spin polarization and spin currents without
the need of magnetic materials.\cite{awschalom,sinovaRMP,hoffmann13}

In particular, at interfaces or surfaces of materials, the breaking
of structure inversion symmetry, combined with the atomic spin-orbit
interaction gives rise to the Rashba (or Bychkov-Rashba) spin-orbit 
coupling.\cite{rashba,winkler} It is well-known that the
Rashba SO coupling (Rashba SOC or RSOC) in two-dimensional
(2D) systems leads to the presence of pure spin currents, which are
transversal to the direction of an applied charge current. This spin 
current is considered the intrinsic mechanism for the spin-Hall effect,
which in many systems coexists and competes with extrinsic mechanisms.
\cite{dyakonov,Hirsch99,murakami06,sinova04,hoffmann13,vignale10}
For finite width 2D systems such as wires or strips, this spin current
is the origin of the phenomenon of spin accumulation.\cite{nikolic,
malshukov} 

The Rashba SOC in noninteracting 2D systems has been widely studied
in first quantization,\cite{sinovaRMP,bercioux} where the finite width 
of strips is enforced by an harmonic potential or by an infinite
potential well.\cite{erlingsson}
Much less has been done using second quantization on lattices, which
provides a road to study the crossover from one-dimensional (1D) to 2D
behavior,\cite{wenk,riera,gothassaad} and to include realistic hoppings
and SO couplings, as well as various types of electron correlations.
These electron correlations are present in a
number of surfaces and interfaces involving transition metal oxides,
such as SrTiO$_3$\cite{hwang,caviglia,benshalom,ykim,caprara},
particularly  LaAlO$_3$/SrTiO$_3$,\cite{liu2014,banerjee,sahin,
liu2014,khalsa,bucheli} and BaTiO$_3$, for example in the
heterostructure BaTiO$_3$/BaOsO$_3$.\cite{seiji15}

\begin{figure}
\includegraphics[width=0.8\columnwidth,angle=0]{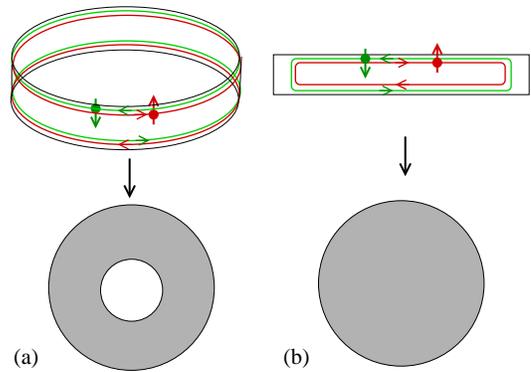}
\caption{(Color online) Strips with (a) periodic boundary
conditions, (b) and open boundary conditions along the
longitudinal direction. In both cases there are open boundary
conditions in the transversal direction. Conventional
helical edge currents are shown for illustration. Each type of boundary 
condition is associated with a topology, as shown in the bottom figures.}
\label{fig1}
\end{figure}

In addition, second quantization and discrete lattices provide a scale
of detail at a microscopic level, not accessible by continuum coarse
grained techniques. The main physical property we want to address
within this approach is the distribution of spin-up and spin-down
electron currents that appear on strips due to the presence of
edges, even in the absence of external electromagnetic potentials or
fields. Helical edge currents where electrons with up and down spins
move in opposite directions, such as those illustrated in 
Fig.~\ref{fig1}, are well-known to be present in topological
insulators of the quantum spin-Hall (QSH) type.\cite{bernevig,sheng,
konig,murakami,kanemele,okudakimura}
In the present work, we examine helical currents which appear
in {\em metallic} planar strips in the presence of Rashba SOC.
These helical currents are slowly decaying from the edges, and hence 
for a narrow strip, depending on the RSOC strength, they are present
over all the section of the strip, not just on its edges as in the
QSH case. This feature should be ultimately related to the
fundamental difference between the Rashba SOC and the SO term that
consists of two copies of the Haldane term and is the essence of the
Kane-Mele model\cite{kanemele} This Haldane-like term of the Kane-Mele
model, contrary to the Rashba SO term, conserves the $z$-component of
the total spin, while breaking the time-reversal symmetry.
We emphasize that what we call in the following Rashba helical
currents (RHC) (defined in section \ref{modelmethod}), are {\em charge}
currents for each spin projection,
and should not be confused with the already mentioned transversal spin
currents, which are the ones related to the spin-Hall conductivity
\cite{sinova04,nomura}. In the absence of appplied magnetic or electric
fields, at any point of the system, the spin-up Rashba helical current
is exactly opposite to the spin-down one, and hence of course there is
no {\em net} charge current.

One of the main features we will study is the texture of these Rashba
induced helical currents, that is, the amplitude and sign oscillations of
spin-up and spin-down electron currents on different chains as a function
of their distance to the edge. These sign oscillations of helical
currents were not so far detected experimentally, and hence did not
receive too much theoretical attention, since in semiconductor devices
the electron densities are low and RSOC is weak. However, some traces
of this texture have been reported theoretically on studies on
semiconductor wires.\cite{nomura2,nikolic}
New materials which are being considered for spintronic devices 
can perform at higher electron densities and present much larger RSOC,
and hence the results of the present study may be relevant for such
systems.\cite{seiji15,giantSO}

Hence, the first purpose of the present article is to numerically
determine the helical currents that appear due to the RSOC in
two different systems, as shown in Fig.~\ref{fig1}. We will first
consider a ring geometry, that is, a strip with open boundary 
conditions (OBC) in the transversal direction and periodic boundary
conditions (PBC) in the longitudinal one (Fig.~\ref{fig1}(a)). Then, we 
will analyze a strip with full OBC (Fig.~\ref{fig1}(b)), which could
be regarded as an idealized version of the wires employed in
spintronic devices.

In Section~\ref{noninteractingcase} we start our study with 
noninteracting systems in the absence of external
electromagnetic sources. By using a combination of momentum and
real-space techniques on both kinds of strips, the Rashba helical
current distribution across the transversal section will be 
determined.

Next, we examine how the Rashba helical currents determine the
net charge currents that appear in the presence of external 
electromagnetic sources. This part of the work, performed in
Section~\ref{currentsfield}, although certainly is the most relevant
to physical systems or devices, could only be properly understood
by resorting to the behavior of the Rashba helical currents
studied in the previous section.  In the case of rings, shown in
Fig.~\ref{fig1}(a), persistent currents can be induced by applying
an Aharonov-Bohm flux through it.  For strips such as those of 
Fig.~\ref{fig1}(b), these currents are induced by connecting the
trips to two leads with a voltage difference. As
shown in Fig.~\ref{fig1}, both kinds of strips belong in 
principle to different topologies and transport physics, and 
the purpose of the present work is to examine how the Rashba helical
currents give rise to net longitudinal currents in both
geometries.

In this second part of our effort, we will also include
the effects of electron correlations, introduced by a Hubbard term
in the Hamiltonian, again relevant for a number of recently 
introduced materials and devices, as mentioned before. So far,
few studies have been accomplished in models including the Rashba
SOC and Hubbard repulsion, and most of them were done on
quasi-onedimensional lattices.\cite{riera,gothassaad}
We will show that the analysis of RHC gives a fresh approach
on this kind of correlated models, particularly on its 
transport properties.

\section{Model and methods}
\label{modelmethod}

The Rashba SO Hamiltonian on the square lattice is given by:
\cite{pareek,sinova14}
\begin{eqnarray}
H_R = V_{SO} \sum_{l} &[&c_{l+x,\downarrow}^\dagger c_{l,\uparrow} -
c_{l+x,\uparrow}^\dagger c_{l,\downarrow} + i (
c_{l+y,\downarrow}^\dagger c_{l,\uparrow} \nonumber  \\
&+& c_{l+y,\uparrow}^\dagger c_{l,\downarrow}) + H. c.]
\label{Rham}
\end{eqnarray}
and the total Hamiltonian is $H=H_H+H_R$, where
\begin{eqnarray}
H_H = - t \sum_{<l,m>,\sigma} (c_{l,\sigma}^\dagger c_{m,\sigma} +
H. c.) + U \sum_{l} n_{l,\uparrow} n_{l,\downarrow}
\label{hamhub}
\end{eqnarray}
corresponds to the Hubbard model. The notation is standard.
We will consider the isotropic case $t_x=t_y=t$, and 
$V_{SO,x}=V_{SO,y}=V_{SO}$, where $x$ ($y$) is the longitudinal
(transversal) directions of the strips shown in Fig.~\ref{fig1},
except otherwise stated.
In the following, we have adopted the normalization,
$\sqrt{t^2+V_{SO}^2}=1$, which we adopt as the unit of energy.

It is well-known that with the Rashba term, the $z$-projection of the
total spin is no longer conserved, as well as the rotational symmetry
in the spin space, while the time-reversal symmetry is still
conserved.

In this work, we will study planar strips, with $L$ sites on the
longitudinal direction, with periodic (Fig.~\ref{fig1}(a)) or open
(Fig.~\ref{fig1}(b)), boundary conditions, and $W$ sites in the 
transversal direction, with OBC. In general, we will take
$L \gg W$, so that the strips could be considered as $W$ coupled
chains. The total number of sites is $N=L \times W$, and the 
total number of electrons, $N_e$, such that the filling is
$n=N_e/N$. We will not consider the presence of impurities in
our system.

The main quantities we are interested in are the currents associated
to the hopping term in Eq.~(\ref{hamhub}), termed the spin-conserving 
current, $J_{\sigma,x,y,\hat{\mu}}$, on each link between $(x,y)$
and $(x,y)+\hat{\mu}$, where $\hat{\mu}=\hat{x}, \hat{y}$, which is
the expectation value of the operator:
\begin{eqnarray}
\hat{j}_{\sigma,x,y,\hat{\mu}} = i t
(c_{(x,y)+\hat{\mu},\sigma}^\dagger c_{(x,y),\sigma} - H. c.),
\label{curhop}
\end{eqnarray}
and the contribution from the RSOC term, which leads to the so-called
spin-flipping current, $J_{SO,x,y,\hat{\mu}}$ which is the expectation 
value of a similar current operator that can be also similarly
derived from Eq.~(\ref{Rham}) as the first order perturbation induced by
a magnetic potential ${\bf A}$ added to Eq.~(\ref{Rham}) by the Peierls
factors, $\gamma_{\hat{\mu}} \rightarrow \gamma_{\hat{\mu}}
\exp{(-i e A_{\hat{\mu}})}$ (in units of $\hbar$), where
$\gamma_{\hat{\mu}}$ are the appropriate coupling constant in the
$\hat{\mu}$-direction. In the following we will also adopt units
where $e=1$.
Let us write down only the $x$-component of the SO current, which
we will need for future discussions:
\begin{eqnarray}
\hat{j}_{SO,x,y,\hat{x}} &=& - i V_{SO}
[(c_{(x,y)+\hat{x},\downarrow}^\dagger c_{(x,y),\uparrow} - H. c.)- 
\nonumber \\
&(&c_{(x,y)+\hat{x},\uparrow}^\dagger c_{(x,y),\downarrow} - H. c.)],
\label{curso}
\end{eqnarray}
The expectation value of each term in parenthesis will be denoted as
$J'_{SO,x,y,\hat{x}}$ and $J''_{SO,x,y,\hat{x}}$, respectively, and
similarly for the $y$-direction.
The total current on each link is then
$J_{tot,x,y,\hat{\mu}}=J_{\uparrow,x,y,\hat{\mu}}+
J_{\downarrow,x,y,\hat{\mu}} +J_{SO,x,y,\hat{\mu}}$.

In the following we will compute and discuss the RHC as the "spin-up 
currents" ("spin-down currents"), defined as the hopping currents
$J_{\uparrow,x,y,\hat{x}}$
($J_{\downarrow,x,y,\hat{x}}$) described above.

Spin accumulation is a well-known consequence of the spin-Hall effect
in the presence of boundaries in the direction transversal to the
direction of the charge current.\cite{nikolic,malshukov,riera}
That is, a charge current in the longitudinal direction that appears
due to an external electromagnetic field, induces a transversal
{\em spin} current, which then leads to spin polarization with opposite
sign on opposite edges of the conducting strip.
In principle, the spin accumulation is defined as the relative 
out-of-plane spin polarization between the two sides of the strip
that is:
\begin{eqnarray}
\Delta S^z=\sum_{1\leq j\leq W/2} S^z(j) -
           \sum_{W/2+1\leq j\leq W} S^z(j)
\end{eqnarray}
where $S^z(j)$ is the sum of the $z$-component of the electron spins
on the chain $j$. For wide strips it is expected that the spin
accumulation is concentrated
on the strip edges as it was observed experimentally~\cite{kato},
although there is no consensus about the interpretation of these
experimental results. In the narrow strips we will consider in 
Section \ref{currentsfield}, the spin polarization could take 
appreciable values across the whole section of the strip, so we will
compute the polarization or spin accumulation on each chain,
$\Delta S^z(j)=S^z(j)-S^z(W-j+1)$, $j=1,\ldots,W/2$. Hence the 
total spin acumulation can be expressed as
$\Delta S^z=\sum_{1\leq j\leq W/2} \Delta S^z(j)$.

It has been noticed that spin polarization could also appear 
on strips due to the inverse spin galvanic effect (ISGE).\cite{wolf}
However, the current-induced spin polarization due to ISGE refers to
{\em in-plane} ($x,y$) components of the spin, and has been studied in 
two-dimensional electron systems for varying $x,y$-directions of the
currents.\cite{ganichev} In contrast, we would like to emphasize that
the spin polarization we compute and discuss in 
Section~\ref{currentsfield}, is the {\em out-of-plane} ($z$)
component related to the Spin-Hall effect on strips,
and hence not to the ISGE.

For the noninteracting case, we solve numerically the Hamiltonian
for a single electron. For the case of rings, Fig.~\ref{fig1}(a),
translation symmetry is preserved along the longitudinal direction
($x$-axis), while it is broken in the transversal direction ($y$-axis)
due to the open BC. In this case, for each $k_x$ one has to solve a
$2 W\times 2W$ matrix, and the single particle eigenvalues and
eigenvectors
will be a function of $(k_x,y,\sigma)$. In the absence of external
electromagnetic fields, static currents on each chain both of the
hopping and SO types, are computed by straightforward ground state
averaging. These zero temperature results are presented in
Section~\ref{noninter.rings}.

For the geometry of Fig.~\ref{fig1}(b), translation symmetry is broken
also in the $x$ direction, so the single particle Hamiltonian matrix is
formulated in real space, and it has dimensions $2 N\times 2N$. In this
case, one has in principle currents of spin-$\sigma$ electrons varying 
on each bond of the lattice, and in both directions, that is
$J_{\sigma,x,y,\hat{\mu}}$. In
Section~\ref{noninter.open}, only results for the currents at the
center of the strip, $J_{\sigma,L/2,y,\hat{\mu}}$
will be discussed.

To deal with both external electromagnetic sources and  nonzero
electron-electron interactions, that is a finite value of $U$ in
the Hubbard term, we will resort to much more involved many-body
techniques to study the zero temperature behavior. For the case
of periodic strips, we will study persistent
currents along the ring, which are induced by piercing the
ring by an  Aharonov-Bohm flux.\cite{splettstoesser,riera,michetti}
The study of these currents
is accomplished by a simple variational Monte Carlo (VMC)
where configurations generated by a trial wave function which
consists of a product of Slater determinants, are weighted
by a Gutzwiller factor which depends on the number of
double-occupied sites and the value of $U$.\cite{honghirsch,
giamarchi}
This technique allows us to study rings up to
$W=4$, and electron-electron Hubbard repulsion up to $U=8$.
Although this simple variational wave function underestimates
the effect of $U$, the obtained results are qualitatively reliable
as compared with those obtained by other techniques. Typically,
for each set of parameters, and for each value of the 
Aharonov-Bohm flux, $\Phi$, we take Monte Carlo averages over
$10^6$ sweeps.

In the case of an Aharonov-Bohm flux $\Phi$ piercing a ring, the total
Hamiltonian $H_R+H_H$ has to be treated in the
presence of Peierls factors in both the hopping and the RSOC terms
along the $x$-direction, and $A_{\hat{x}}$ should be replaced by
$\Phi/L$, where in the adopted units, $0 \leq \Phi \leq 2 \pi$. In
this case, the ground state eigenvalue and eigenvector will be
$E_0=E_0(\Phi)$, $\Psi_0=\Psi_0(\Phi)$, and hence the ground state
expectation value of any physical quantity will also depend on $\Phi$.

To study currents induced by an external electric field, as well as
electron correlations, in strips where OBC has been imposed also in the
longitudinal direction, we resort to density matrix-renormalization
group (DMRG) technique,\cite{schollwock} which is in principle exact
for static properties. In this case, currents appear as a time
dependent feature after applying a voltage bias between the ends
of the strip at a given time, after the iterative process has
converged to the ground state. This first standard DMRG was carried 
out keeping up 800 states, and performing at least 14 sweeps, enough 
to achieve truncation errors of the order $O(10^{-6})$.
The voltage bias is applied between the two halves of the strip in
the longitudinal direction. Alternatively, we have
applied this bias to the two ends of the strip without 
affecting the two central columns of the strip. The results 
obtained with both recipes are virtually identical. The voltage
bias is small, $\Delta V=0.01$, so as to avoid strong
nonequilibrium processes.  This calculation requires an 
extension of the original DMRG algorithm, and the results are
considered "quasi-exact".
Among different DMRG approaches to deal with time evolution
problems\cite{schollwock,feiginwhite,alhassanieh}, we choose the
algorithm introduced by Manmanna {\it et al}\cite{MWNM-09}, which
is a variant based on a Krylov-space representation of the time
evolution operator\cite{schmitteckert}. Thereby, we are able to evolve
systems with interactions involving many coupled chains, and obtain
very accurate results for long time periods. We observed that using
at each time slice $dt=0.1$, an Krylov base of $\sim 30$ states, the
errors become smaller than the symbol sizes. For each strip, and for
a given set of parameters, a single value for each type of current
was taken as the amplitude of the first maximum reached in its
time evolution, considered as an approximation of the expected
plateau.

The application of computational techniques, particularly DMRG,
is severely limited by the non-conservation of the total $S^z$,
which implies a much larger Hilbert space for the system. The need
to work with complex numbers, also implied by the Rashba SO coupling,
further increases CPU time and memory requirements.

\section{Noninteracting case}
\label{noninteractingcase}

\subsection{Rashba helical currents}
\label{handwaving}

Let us start with the noninteracting case, $U=0$. As we will
extensively show in the next sections, the RHC consist of spin-up or
spin-down hopping currents which are exactly opposite on each chain
across the strip section in the absence of external electromagnetic
sources (fields or potentials). In addition, the currents for a
given spin orientation, have opposite
direction on chains symmetrically located with respect to the middle
of the section, and their sign and strength vary over the section.
In principle, counter propagating currents for each spin projection
are guessed from general arguments similar to those leading to the
prediction of the spin-Hall effect\cite{sinova04,okudakimura} and
which ultimately stem from requirements of time-reversal symmetry.

\begin{figure}[t]
\includegraphics[width=0.8\columnwidth,angle=0]{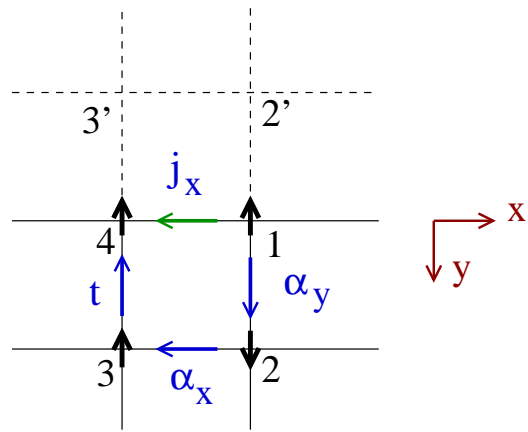}
\caption{[Color Online] Schematic explanation of the origin of an
effective current between sites '1' and '4', located at the 
system edge. $\alpha_x$, $\alpha_y$ and 't', indicate a Rashba SO
movement along $x$ and $y$ directions, and a regular hopping 
respectively. Dashed lines indicate the possible processes that
would be present in the bulk.}
\label{figscheme}
\end{figure}

An alternative and instructive way of understanding the presence of these
currents with opposite circulation of spin up and down electrons in the
absence of external sources, can be done in real space in systems with
Rashba SOC. At an effective level,
these Rashba helical currents appear due a term in the Hamiltonian,
\begin{align}
H_{\hat{\mu}}^{eff} &\sim& E_{\hat{\mu}} j_{\hat{\mu}}~~~~~~~~~~~~~~~~~~~~~~~~~~~~~~~~~~~~
~~~~~~~~~~~~~  \nonumber \\
  &\sim& i \lambda \sum_{l} [(c_{l+\hat{\mu},\uparrow}^\dagger
 c_{l+\hat{\mu},\downarrow}^\dagger) (i\sigma^z) \left( \begin{array}{c}
  c_{l,\uparrow}  \\
  c_{l,\downarrow}
  \end{array} \right) - H. c.]~~~
\label{effedg}
\end{align}
where $\hat{\mu}=\hat{x}$ or $\hat{y}$.
The origin of this term is schematically shown in Fig.~\ref{figscheme}.
A spin up electron at site "1" could reach site "4" by moving to site "2"
involving a spin-orbit process, and then to site "3" with another SO
process. Finally, it could reach site "4" by an ordinary hopping.
The full process would involve $\sigma^y \times \sigma^x =-i \sigma^z$,
with a factor $\lambda$ proportional to $t V_{SO}^2$.
An analogous process along sites "2'~" and "3'~" would have an opposite
sign but this contribution is absent if that line is on a system edge. 
It is easy to see that the reverse process from "4" to "1" would lead
to $\sigma^x \times \sigma^y =i \sigma^z$, thus providing the sign
for the current in Eq.~\ref{effedg}. In addition, processes involving
more sites further away from the edge would contribute to that 
effective current on the edge, and also on legs located in the
bulk due to the lack of compensation.  If sites "1" and "4" belong
to a leg located at $y$, processes through sites "2" and "3",
located at "y+a" and process through sites "2'~" and "3'~", located
at "y-a" would not cancel each other because legs "y+a" and "y-a" are
in general not symmetric due to the proximity to the edge. For
example, an effective current involving five sites
would lead to a current between sites at the line next to the edge,
and so forth. Of course, each hopping would on average have a
probability of $(1-n/2)$, where $n$ is the electron filling, and
assuming an equal number of electrons with up and down spins.
Notice that for the so-called persistent helix state,\cite{bernevig2}
where the Rashba and Dresselhaus,\cite{dresselhaus} couplings
are present with equal strength, the SO coupling is mediated by 
$\sigma^x - \sigma^y$  in both directions, and hence the effective 
terms illustrated in Fig.~\ref{figscheme} turns out to be of the
kinetic type, not of the current type.

Of course, for very large strip widths, processes with longer paths become
unlikely, and far from the edges, processes with shorter path would
cancel because legs become approximately equivalent. It is expected then
that for large strip widths, these currents become essentially nonzero
only at the strip edges. It will be tempting to assign the changes in
sign and strength as a function of the distance of the chains to the
edge, to a sort of Friedel oscillations, as it will be explored below.

It should also be noticed that the Rashba Hamiltonian Eq.~(\ref{Rham}) 
can be written in terms of currents, and the same kind of currents
here studied could be the one seen in Ref.~\onlinecite{bovenzi} in
the presence of impurities. That is, the impurities provide the 
breaking of translational symmetry that appear in our case due to
the open boundaries of strip geometry.

Let us remark that according to the previous arguments, the RHC are of
the hopping or spin-conserving type, given by Eq.~(\ref{curhop}), not of
the SO or spin-flipping type, corresponding to Eq.~(\ref{curso}).

\begin{figure}[t]
\includegraphics[width=0.75\columnwidth,angle=0]{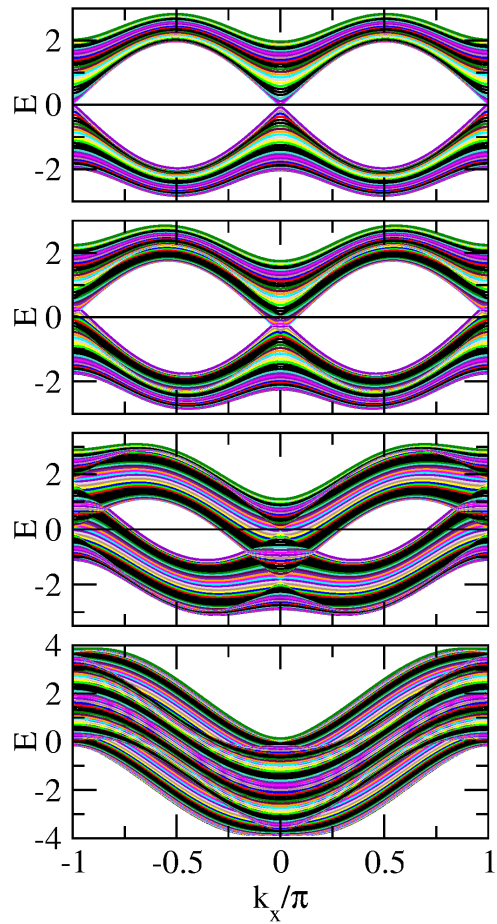}
\caption{[Color Online] Energy bands as a function $k_x$
from top to bottom $V_{SO}/t=64$, 8.0, 2.0, and 0.4.
$64 \times 2000$ ring.}
\label{bandsring64}
\end{figure}

\begin{figure}[t]
\includegraphics[width=0.8\columnwidth,angle=0]{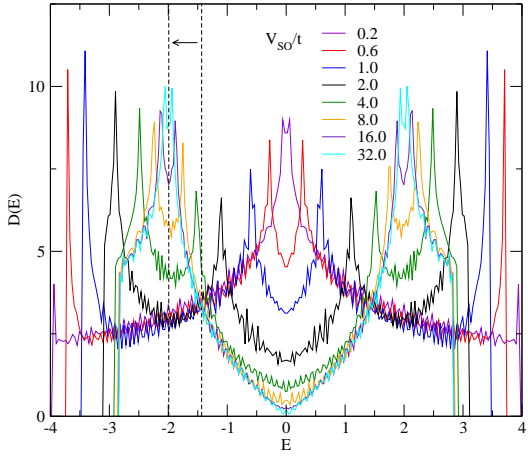}
\caption{[Color Online] Density of states for the $64 \times 2000$
ring for various values of $V_{SO}/t=64$, as indicated on the plot.
The vertical dashed lines indicate the corresponding shift of the
chemical potential for quarter filling.}
\label{dos64}
\end{figure}

In Fig.~\ref{bandsring64} the evolution of energy bands as a function
of  $k_x$ for various values of $V_{SO}/t$ is shown for rings. Results
were obtained for the $64 \times 2000$ ring. The top panel, 
corresponding to $V_{SO}/t\gg 1$, resembles the typical situation found
in topological Dirac semimetals\cite{mzhasan14,sunzhu,baum2015},
with its characteristic Dirac nodes in the Fermi
surface.\cite{dzero,ynagaosa} However, the purpose of the present work
is to study the evolution of Rashba helical currents when $V_{SO}/t < 1$,
when the electron density is reduced below half-filling, and hence the
physics of the topological Dirac semimetals is not relevant.
As $V_{SO}/t$ is
reduced, a more conventional band metallic behavior starts to appear. At
$V_{SO}/t=0.4$, a single conduction band appears. It is interesting to
note that the behavior shown in Fig.~\ref{bandsring64} is also similar
to the one found in the ferromagnetic
Kondo lattice model with Rashba SOC in 2D, with classical localized
spins, for large values of the Hund coupling at quarter 
filling.\cite{meza}
Of course, this band structure is different to the one characteristic
of QSH systems, reflecting the insulating character of the bulk, and
the metallic character of the edge.\cite{kanemele,laubach}

\begin{figure}[ht]
\vspace{0.8cm}
\includegraphics[width=0.9\columnwidth,angle=0]{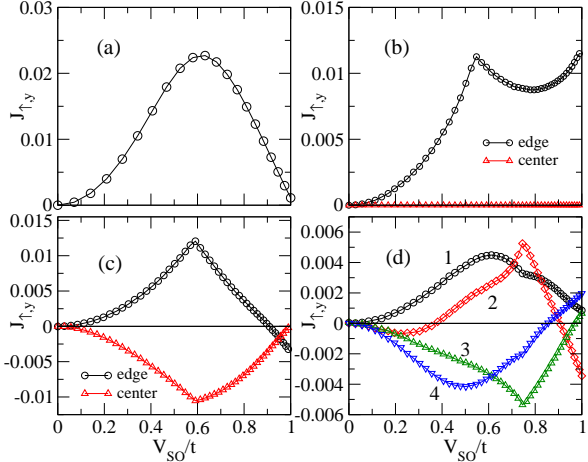}
\caption{[Color Online] Spin-up currents for rings with (a) 2, (b) 3,
(c) 4, and (d) 8 legs as a function of $V_{SO}/t$, and for different
legs as indicated in the plots, $L=8000$, $n=0.5$. In (d), leg "1"
("4") corresponds to the edge (center).}
\label{figvso23}
\end{figure}

The density of states for the same $64 \times 2000$ ring in the 
noninteracting case is shown in Fig.~\ref{dos64}, for various values
of $V_{SO}/t$. It can be clearly seen a suppression of energy states
around zero energy, corresponding to half-filling, as $V_{SO}/t$ is
increased from 0.2 to 64. The vertical dashed lines indicate the 
corresponding shift of the chemical potential for quarter filling.
Both Figs.~\ref{bandsring64} and \ref{dos64} indicate a clear 
metallic behavior for all $V_{SO}/t$ and electron fillings.

\subsection{Results on finite width rings}
\label{noninter.rings}

In this Subsection, we study planar rings such as the one depicted in
Fig.~\ref{fig1}(a).
Since as said earlier, Rashba helical currents of spin-down electrons
are exactly opposite to the spin-up ones, in the following we will
only consider the spin-up electron currents on each chain. Moreover, 
since currents are spatially antisymmetric on chains with respect 
to the central longitudinal axis, we will only show the currents
on the chains in half the section of the strip. Finally, due to
translation invariance along the longitudinal direction, the currents
will not depend on $x$. Hence, in all figures
below, we will depict $J_{\uparrow,l}$, where $l=1,\ldots, W/2$ or
$J_{\uparrow,\nu}$, where $\nu$ is the normalized distance of the chain
to the edge, or chain depth, defined in such a way that $\nu =0$ at the
edge and $\nu =1$ at the center, independently of the width $W$.

As discussed in Subsection~\ref{handwaving}, the RHC, computed here as
$J_{\uparrow,l}$, is a spin-conserving current, and spin-flipping currents
are zero everywhere, which is remarkable since the SO coupling is needed
to be nonzero in {\em both} directions for the RHC to exist. Moreover,
notice that the conservation of the average charge on each site in the
stationary regime (Kirchhoff's law) is trivially satisfied for up and
down spin electrons separately. This is compatible not only with the
net SO currents being zero but also with the two contributions 
$J'_{SO,x,y,\hat{\mu}}$ and $J''_{SO,x,y,\hat{\mu}}$, possibly being zero
in both directions, $\hat{\mu}=\hat{x},\hat{y}$.

\begin{figure}[t]
\includegraphics[width=0.85\columnwidth,angle=0]{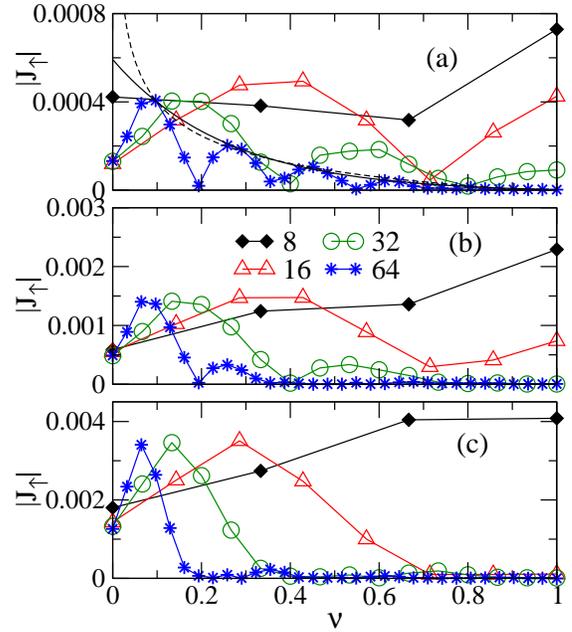}
\caption{[Color Online] Absolute value of spin-up currents on each chain
as a function of the depth of the chain ($\nu=0$, edge, $\nu=1$, center 
leg), (a) $V_{SO}/t=0.1$, (b) 0.2, (c) 0.4. $L=8000$ rings,
$n=0.5$. Widths of the rings are indicated in the
plot. The full (dashed) curves added in (a) correspond to 
exponential (power law) fits.}
\label{figvslegmod}
\end{figure}

\begin{figure}[ht]
\includegraphics[width=0.85\columnwidth,angle=0]{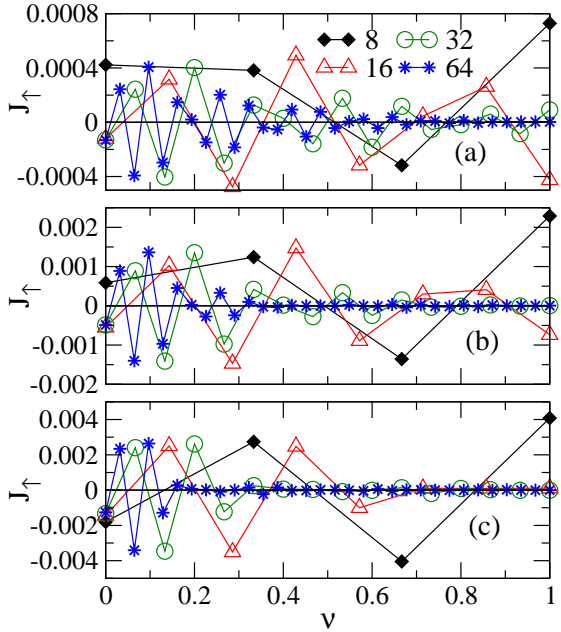}
\caption{[Color Online] Spin-up currents on each chain as a function
of the normalized depth of the chain for (a) $V_{SO}/t=0.1$, (b)
0.2, (c) 0.4. $L=8000$ rings, $W$ indicated
on the plot, $n=0.5$.}
\label{figvsleg}
\vspace{-0.1cm}
\end{figure}

\begin{figure}[t]
\includegraphics[width=0.85\columnwidth,angle=0]{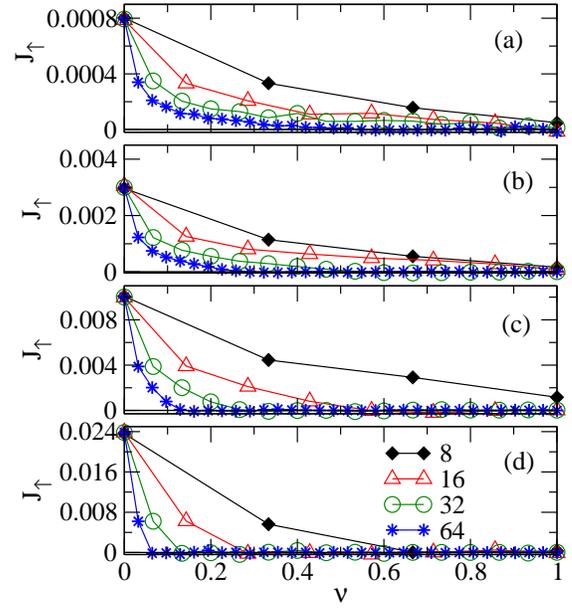}
\caption{[Color Online] Spin-up current on each chain as a function of 
the depth of the chain ($\nu=0$, edge, $\nu=1$, center leg),
for (a) $V_{SO}/t=0.1$, (b) 0.2, (b) 0.4, and (c) 0.8. $L=8000$
rings, $n=1$. Widths of the rings are indicated in the plot.}
\label{figvslegn1}
\vspace{-0.3cm}
\end{figure}

A direct evidence of the existence of the RHC and their predicted
dependence with the square of $V_{SO}/t$ can be observed for strips
of $W=2,3,4$ and 8 at $n=0.5$ at small $V_{SO}/t$, as shown in
Fig.~\ref{figvso23}. This dependence becomes increasingly complex as
the strip width is increased. For all $W$, there is a clear change
of behavior around $V_{SO}/t \approx 0.5$, and for $W >2$, this 
behavior presents some cusps, which are typical of level crossings.
In this, and in the following calculations in this Subsection, 
results for $L=800$ are virtuable indiscernible from those for
$L=8000$, hence we are confident that the results are representative
of the continuum limit in the longitudinal direction.

As suggested by the arguments of Subsection~\ref{handwaving}, the RHC
should concentrate near the strip edges, and only due to its slow
decay from the edges, for small widths, the RHC appears to pervade the
whole system. To examine this behavior we will consider a broad range
of strip widths. As it can be
seen in Fig.~\ref{figvslegmod}, the module of $J_{\uparrow}$ as a 
function of the chain's depth $\nu$ indicates that only for
$W\geq 16$ and $V_{SO}/t\geq 0.4$ it is apparent that these
currents become essentially different from zero near to the strip
edges. In general, wider strips have to be considered to show an
edge-like behavior for smaller SO couplings. Two more features
become apparent. First, consistently with the predictions of
Subsection~\ref{handwaving}, and with the behavior already shown
in Fig.~\ref{figvso23}, the amplitude of these currents increases
almost quadratically with $V_{SO}/t$, and second, and more 
interesting, $J_{\uparrow,\nu}$ presents an oscillatory behavior as a
function of the chain depth, reminiscent of a Friedel oscillation
behavior.

To quantify the decay of RHC as the chains are located away from the
edge, we have fitted the curves for the widest ring, with $W=64$,
with an exponential law, $J_{\uparrow,\nu}\sim \exp{(-B\nu)}$, and 
by a power law, $J_{\uparrow,\nu}\sim \nu^{(-B)}$. Results from these
fits for $V_{SO}/t=0.1$ are shown in Fig.~\ref{figvslegmod}(a),
and a systematic study of this fitting will be presented below in
Fig.~\ref{fitdecay}.

In Fig.~\ref{figvsleg}, it can be noticed that the amplitude
oscillation observed in the previous Figure, actually comes from
a sign oscillation. In fact, close to the edge, and more clearly
for larger widths,  it seems that the oscillations would have
a wavenumber $\pi$, for these results at quarter filling
($n=0.5$).

Let us now consider the case of half-filling ($n=1$). The Rashba
helical currents on each chain are shown in Fig.~\ref{figvslegn1}
for the same values of $W$ and $V_{SO}/t$ as before, now including
also $V_{SO}/t=0.8$. Similarly to the behavior observed for $n=0.5$,
the RHC tend to concentrate near the strip edges as $W$ and $V_{SO}/t$
increase. As a difference with the case of $n=0.5$, at $n=1$
the RHC is maximal at the edges, smoothly decaying
towards the strip center. The absence of sign and amplitude
oscillations for this density $n=1$, is, as in the $n=0.5$ case,
compatible with a wavenumber $2 n \pi$ ($n=1$ at half-filling), which
would correspond to $4k_F$ in a quasi-onedimensional system.

In Fig.~\ref{fitdecay} the coefficients $B$ of the above mentioned
exponential and power law fits of the RHC for (a)  $n=0.5$ and
(b) $n=1$, are shown. The results in (a) were obtained from the
values of $J_{\uparrow,\nu}$ shown in Fig.~\ref{figvslegmod}, and
the results in (b) were obtained form the corresponding values
in Fig.~\ref{figvslegn1}. In both cases, the largest ring width,
$W=64$ was considered. Although these results have somewhat
large error bars, estimated by adopting different subsets of 
points on each curve, it is clear in both cases that $B$ grows
as a function of $V_{SO,x}/t$, indicating a faster decay from the
edges. For $n=0.5$, and the values $V_{SO,x}/t$ considered, the
quality of the fittings using both laws, as measured by the 
correlation coefficient, is similar, although in principle a
power law behavior would seem more natural for a gapless 
system. On the other hand, for $n=1$, the exponential fit is
definitely poor, and only the power law fit leads to minimally
acceptable correlation coefficient.

\begin{figure}[h]
\vspace{1.4cm}
\includegraphics[width=0.88\columnwidth,angle=0]{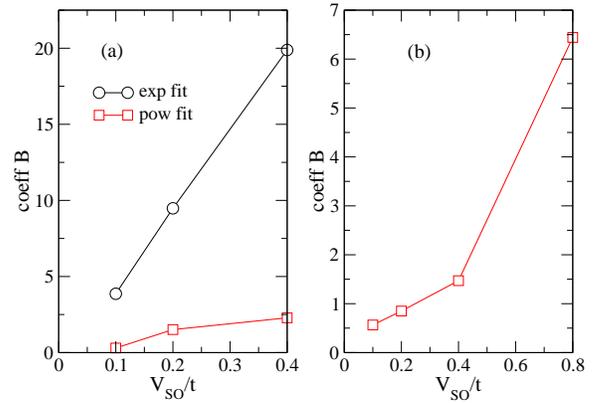}
\caption{[Color Online] Coefficient $B$ of the exponential and
power law fits (see text) of the RHC for (a) $n=0.5$ and
(b) $n=1$, as a function of $V_{SO,x}/t$. Error bars are at least
twice the size of the symbols used.}
\label{fitdecay}
\end{figure}

\begin{figure}[ht]
\vspace{0.1cm}
\includegraphics[width=0.9\columnwidth,angle=0]{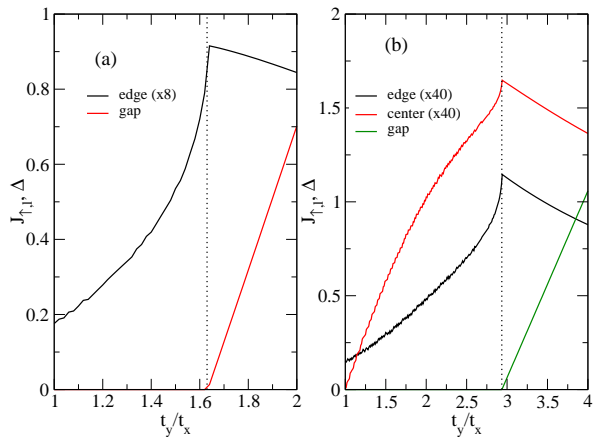}
\caption{[Color Online] Spin-up currents on each chain as a 
function of $t_{y}/t_x$ for (a) 2-leg and (b) 4-leg rings,
$L=800$, $V_{SO,x}/t_x=V_{SO,y}/t_y=1$, $t_x=1$, $n=0.5$.}
\label{figgap24}
\end{figure}

\begin{figure}[t]
\includegraphics[width=0.78\columnwidth,angle=0]{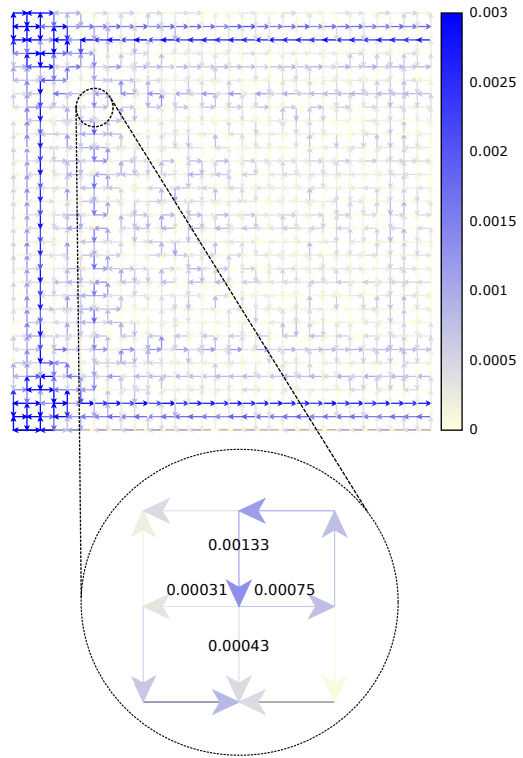}
\vspace{-1.8cm}
\caption{[Color Online] Spin-up currents on each nearest neighbor link,
near the left end of the $256 \times 32$ strip with OBC, $V_{SO}/t=0.8$
$n=0.5$. The colour of each arrow is proportional to the current
strength, according to the included colour code.
In the zoom, the currents on the links merging at the indicated site,
are shown together with their absolute values.}
\label{opencurrents}
\end{figure}

\begin{figure}[ht]
\includegraphics[width=0.8\columnwidth,angle=0]{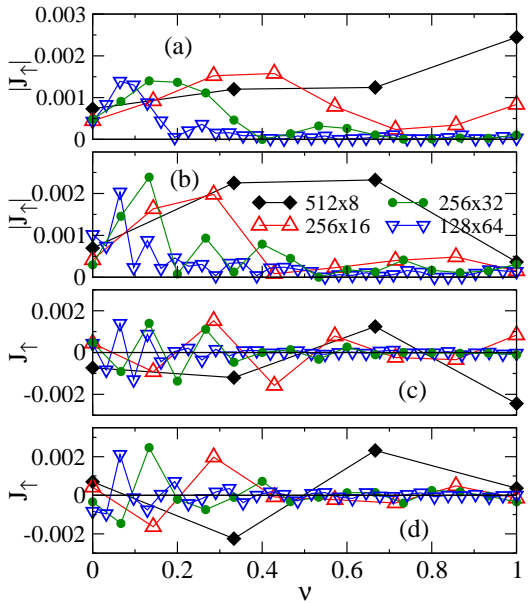}
\caption{[Color Online] Absolute value of spin-up currents at the
chain center as a function of the depth of the chain, (a)
$V_{SO}/t=0.2$, (b) 0.8. Spin-up currents at the
chain center, (c) $V_{SO}/t=0.2$, (d) 0.8.
$L\times W$ indicated on the plot. Strips with OBC, $n=0.5$.}
\label{obcvslegn05mod}
\end{figure}

Let us emphasize that the behavior discussed in this Section
occurs in the metallic phase. In Section~\ref{voltdmrg}, the effect
of the Hubbard term and a possible Mott-Hubbard type of 
metal-insulator transition (MIT) will be considered. 
However, before leaving this Subsection, we can explore the
effect of a MIT due to the opening of a band gap which may occur
for sufficiently large ratio $t_y/t_x$, and at specific
electron fillings. To this end, we should consider an spatially
anisotropic extension of the Hamiltonian, where we keep the ratio
$V_{SO,x}/t_x=V_{SO,y}/t_y=1$, and we do not use the normalization
defined in Section~\ref{modelmethod}, but we used $t_x=1$ instead.
In Fig.~\ref{figgap24}, a change of behavior in the Rashba helical
currents is observed when a band gap opens in strips of widths
$W=2$ and 4, at $n=0.5$. It is apparent that in the gaped region
the RHC decrease approximately
following a linear dependence, but clearly a more systematic
research with more realistic band structures should be performed.

\subsection{Results on open strips}
\label{noninter.open}

We now turn to the geometry depicted in Fig.~\ref{fig1}(b), with open
boundary conditions in both directions. In this case, as shown in that
Figure, the RHC would follow closed circuits. However, the currents on
vertical
segments would also present oscillatory behaviors similar to those
shown previously for the currents on rings in the longitudinal
direction. The result of the interference between horizontal and vertical
oscillations leads to complex patterns, as the one illustrated in
Fig.~\ref{opencurrents}, that are little resembling the closed loops
illustrated in Fig.~\ref{fig1}(b), although some segments of closed
loops are still visible near the edges. 

One important difference between the full open strips considered in
this Subsection, and rings, studied in the preceding one, is 
that spin-up and spin-down electron densities at each site, are no
longer conserved separately, that is, the sum over the links
connected to a given site of the hopping currents for each spin
is no longer zero. This can be checked for example with the values
of the currents in the zoom of Fig.~\ref{opencurrents}. However,
the Kirchhoff law for the {\em total} electron density at each site
is still satisfied since on each link, 
$J_{\uparrow,x,y,\hat{\mu}}=-J_{\downarrow,x,y,\hat{\mu}}$, 
and the net SO current on each link is zero. It is interesting to
notice, that in order to conserve the density of up electrons on
each site, the currents $J'_{SO,x,y,\hat{x}}$ defined after 
Eq.~(\ref{curso}) must be nonzero in such a way that its sum over
the links connected to a site must be equal to the corresponding
sum of spin-up hopping currents. A similar conclusion can be
reached for the sum of the spin-down currents and the other term of
the SO currents, $J''_{SO,x,y,\hat{x}}$.

From the above discussion, to study the dependence of
RHC on chains as a function of their distance to the edge, we have
considered for simplicity the currents on the central link of these
chains, that is $J_{\uparrow,L/2,\nu}$, where we have always adopted
$L$ even. To reduce finite size effects, for each width we have taken
the largest value of $L$ accessible to our computational facilities.

Figs.~\ref{obcvslegn05mod}(a),(b) show results for the absolute value
of $J_{\uparrow,L/2,\nu}$, at quarter filling, for $V_{SO}/t=0.2$ and
0.8, and for various strip widths. The overall behavior is similar
to that found previously for rings for the same sets of parameters,
that is, RHC is concentrated close to the edge for large $V_{SO}/t$
and $W$, while for small RSOC strength and widths, the overlapping of
the tails of the decay from teh edges leads to RHC finite over the
whole strip section. Results for $W=64$ are probably
affected by finite size effects on the length $L$.

Similarly also to the case of rings, the oscillations actually stem
from sign oscillations, as it can observed in
Figs.~\ref{obcvslegn05mod}(c),(d), for the same parameters as in
Figs.~\ref{obcvslegn05mod}(a),(b), respectively.
Again, it seems that this oscillation has a wavenumber of $\pi$ at
this density. Overall, the results for $J_{\uparrow,L/2,\nu}$ seem more
erratic than the results for $J_{\uparrow,\nu}$ considered before for
rings: although the later is a global quantity, the former is a
local one.

\begin{figure}[t]
\includegraphics[width=0.8\columnwidth,angle=0]{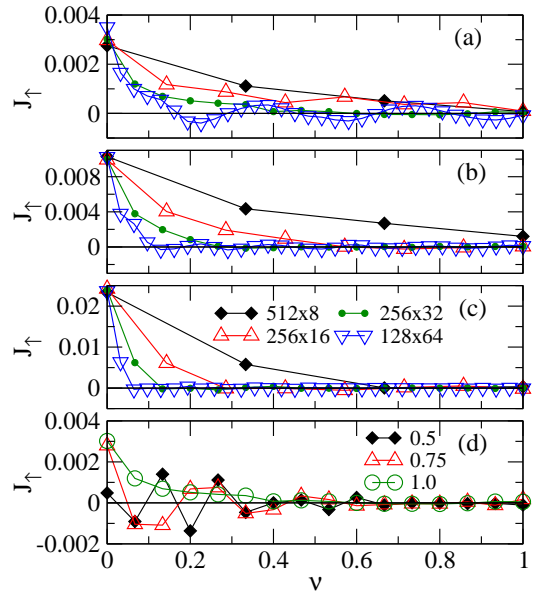}
\caption{[Color Online] Spin-up current at the center of the
chain as a function of the chain position, for (a) $V_{SO}/t=0.2$,
(b) 0.4, (c) 0.8, (d) 0.8, and various values of the strip size,
$L\times W$, as indicated on the plot, $n=1$. In (d) $J_{\uparrow}$
is shown for $V_{SO}/t=0.2$, and various values of the electron
filling indicated on the plot, $256\times 32$.
Strips with OBC.}
\label{curobcn1}
\end{figure}

In addition, a similar dependence of the results with the electron
filling can be inferred by observing the results depicted in
Fig.~\ref{curobcn1}, corresponding to half-filling. Effects of
$V_{SO}/t$ and $W$ are shown in Fig.~\ref{curobcn1}(a)-(c),
with an overall behavior similar to that shown before for rings,
Fig.~\ref{figvslegn1}, that is, an apparent absence of oscillations
except for the $128\times 64$, probably due to finite size effects.
A somewhat more systematic study of the oscillations
dependence on the electron filling is presented in 
Fig.~\ref{curobcn1}(d), where results for $V_{SO}/t=0.2$, on the
$256\times 32$ strip, for the fillings $n=0.5$, 0.75 and 1. For
$n=0.75$, it seems that the oscillation has period four, consistent
with a $4k_F$ value of $\pi/2$ in the first Brillouin zone.

To close this Subsection, let us emphasize that although it is not
possible for open strips to reach the continuum limit as it was
chieved in rings, the fact that one obtains essentially the same
behavior indicates that finite size effects, and in lattice systems,
the consequent discretized energy spectrum, are neligible.

\section{Interacting case: currents due to external electromagnetic
sources}
\label{currentsfield}

\begin{figure}[t]
\includegraphics[width=0.9\columnwidth,angle=0]{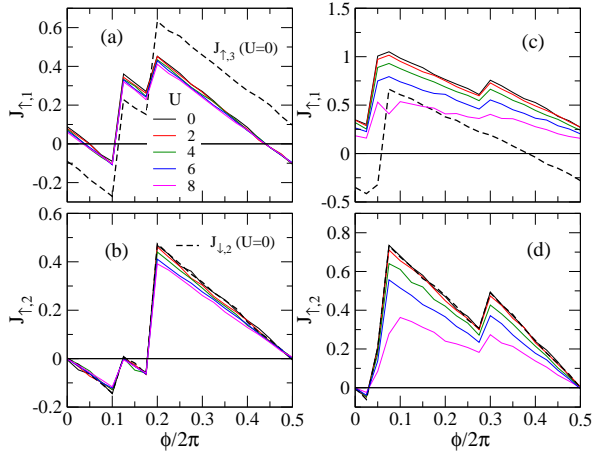}
\caption{[Color Online] Current of spin-up electrons as a function of 
the AB flux and various values of $U$ indicated on the plot, for
$V_{SO}/t=0.6$, (a) edge chains $y=1$, $n=0.5$, (b) center chain,
$n=0.5$, (c) edge chain
$y=1$, $n=1$, (d) center chain, $n=1$. In (a) and (c),
$J_{\uparrow,3}$ for $U=0$ was included. In (b) and (d),
the spin-down current for $U=0$ has been added for comparison.
Results obtained with VMC on the $16\times 3$ ring.}
\label{curlegflux}
\end{figure}

In this section, we will not only address the evolution of net
carrier currents out of helical currents in Rashba metallic
strips when external electromagnetic fields are applied, but also we
will examine how electron correlations in the form of a Hubbard
repulsion could affect those Rashba helical currents.

\subsection{Persistent currents}
\label{ringsVMC}

We will consider first how the RHC in the presence of persistent
currents in rings will lead to net charge currents on each chain, 
both of the spin-conserving and the spin-flipping types, and to
polarized currents, as well as spin accumulation.
Persistent currents induced by an Aharonov-Bohm (AB) flux, are 
studied in this section by VMC, up to $U=8$.
The value of each physical quantity obtained for a given $V_{SO}/t$
and $U$, density, and lattice size, is adopted as the average over
the AB flux between 0 and $\pi$. Of course other
criterions could be adopted, for example to take the maximum of the
absolute value of each quantity on that interval of the AB flux.

The applied AB flux installs a net persistent hopping current,
$J_{hop}/2=\sum_l J_{\uparrow,l,\hat{x}}=
\sum_l J_{\downarrow,l,\hat{x}} \neq 0$, where the sum extends over
all the chains of the strip. Since there is a net charge
current along the longitudinal direction, it is well-known that
a torque will appear on the spins making them rotate in the
$(x,z)$ plane\cite{okudakimura}, that is, there will be a net
SO or spin-flipping current along the $x$-direction, $J_{SO}$.
Then, the total current in the $x$-direction is
$J_{tot}=J_{hop}+J_{SO}$

Let us start by discussing the case of three-chain rings. In 
Fig.~\ref{curlegflux}, the evolution of RHC currents on each chain, 
$J_{\uparrow,y}$, with the AB flux, $\Phi$ is shown for $V_{SO}/t=0.6$,
on the $16\times 3$ ring.
In this, and in similar plots, we display properties as a function
of the AB flux in the range
$0 \leq \Phi \leq \pi$, taking advantage of the symmetry
$O(\Phi)=-O(2\pi-\Phi)$, where $O$ is any physical quantity.
The relation involving the outer chains,
$J_{\uparrow,1}=-J_{\uparrow,3}$, is broken as soon as $\Phi$ is different
from zero, as illustrated in Fig.~\ref{curlegflux}(a), for $n=0.5$ and
$U=0$.
However, the conditions $J_{\downarrow,1}=J_{\uparrow,3}$ and
$J_{\uparrow,1}=J_{\downarrow,3}$, still hold for any $\Phi$. On the inner
chain, it is observed, for $n=0.5$, (Fig.~\ref{curlegflux}(b)) that
$J_{\downarrow,2}=J_{\uparrow,2}$ for all $\Phi$. The same relations are
present at half-filling,  $n=1$, as shown in Fig.~\ref{curlegflux}(c)
for the outer chains, and in Fig.~\ref{curlegflux}(d) for the inner
chain. The dependence of $J_{\uparrow,y}$ with the Hubbard repulsion $U$ is
also included in this plot. One should remember that in general VMC
underestimates the effect of $U$. Nonetheless, it is quite apparent
that the effects of $U$ at $n=1$ are stronger than at $n=0.5$.
In fact, as we will see in the next Subsection, where the DMRG
technique exactly treats the electron correlations, the model
considered experiences a Mott-Hubbard type of metal-insulating
transition, given by the vanishing of the total currents.

\begin{figure}[t]
\includegraphics[width=0.9\columnwidth,angle=0]{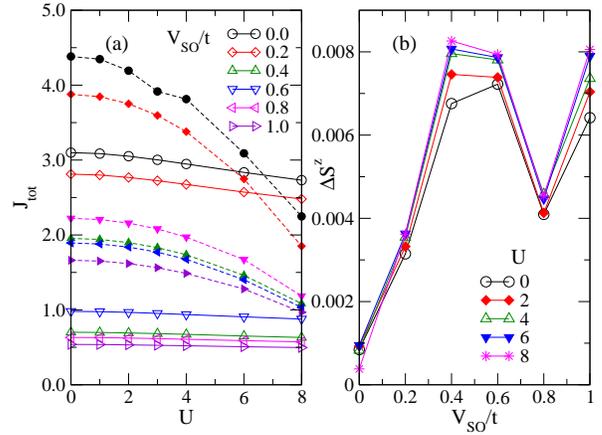}
\caption{[Color Online] 
(a) Total current as a function of $U$ and
for various $V_{SO}/t$ as indicated on the plot. Results for
$n=0.5$ ($n=1$) are shown with empty (full) symbols. (b)
Total spin accumulation as a function of $V_{SO}/t$ and various
values of  $U$ as indicated on the plot, $n=0.5$.
Results obtained with VMC on the $16\times 3$ ring.}
\label{avesumL3}
\end{figure}

The results of a systematic study as a function of both
$V_{SO}/t$ and $U$, are depicted in Fig.~\ref{avesumL3}. In 
Fig.~\ref{avesumL3}(a) it can be observed that the flux-averaged
total currents decrease as a function of both $V_{SO}/t$ and $U$.
For a fixed $V_{SO}/t$, this decreasing of $J_{tot}$ as a 
function of $U$ is much stronger for $n=1$ than for $n=0.5$, 
as expected on general grounds for the Hubbard repulsion.
For both densities, and for any fixed value of $U$, the decrease of
$J_{tot}$ with $V_{SO}/t$ is in general more important than the one
due to $U$. In particular at $U=0$, this result for $W=3$ seems in
contradiction with some suggestion in the literature that the
longitudinal conductivity is independent of $V_{SO}/t$ and
density.\cite{wong10} Further numerical support for the observed
variation of
$J_{tot}$ with $V_{SO}/t$ and density will be given in the next
Subsection, and further discussion will be provided in the final
Section. Results for the flux-averaged 
spin accumulation, is shown in Fig.~\ref{avesumL3}(b). There is
a general trend that $\Delta S^z$ reaches its maximum between
$V_{SO}/t=0.4$ and 0.6 (the result for $V_{SO}/t=1.0$ is a finite size
effect, since it is expected that $\Delta S^z$ vanishes as 
$V_{SO}/t \rightarrow \infty$), and more importantly, there is a
general trend towards
an enhancement of the spin accumulation with the Hubbard coupling
$U$. Both features were already observed for the $W=2$ 
strip\cite{riera}, and together with the results provided below
for the $W=4$ strip, suggest that they could be 
valid for general strip widths $W$.

\begin{figure}[t]
\includegraphics[width=0.9\columnwidth,angle=0]{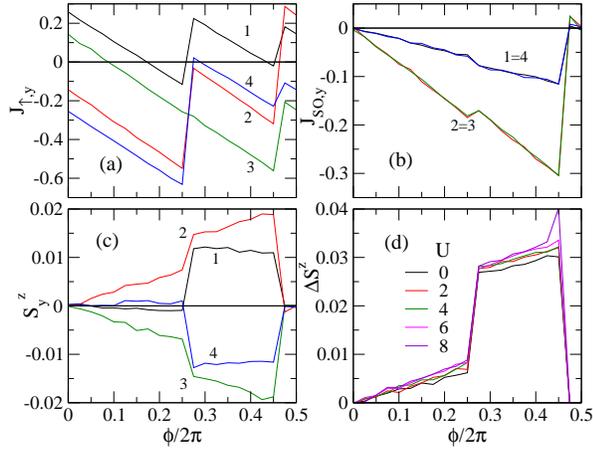}
\caption{[Color Online] (a) Current of spin-up electrons, (b) SO
current, and (c) total $S^z$, as a function of the AB flux $\Phi$,
$U=0$. (d) Total spin accumulation as a function of the AB flux
$\Phi$, and various values of $U$ indicated on the plot.
Results obtained with VMC on the $16\times 4$ ring, $n=0.5$,
$V_{SO}/t=0.6$.}
\label{curL4leg}
\end{figure}

\begin{figure}[t]
\includegraphics[width=0.9\columnwidth,angle=0]{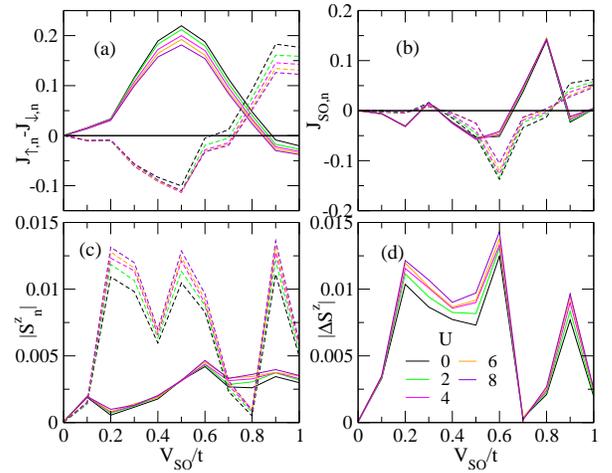}
\caption{[Color Online] (a) Difference between spin-up currents
(see text),
(b) SO currents, and (c) absolute value of $\Delta S^z$, on external
(full lines) and internal (dashed lines) chains, labelled on
the plots, as a function of
$V_{SO}/t$, and various values of $U$. (d) Total spin
accumulation (in absolute value) for various values of $U$
indicated on the plot.
Results obtained with VMC on the $16\times 4$ ring, $n=0.5$.}
\label{all4lsoU}
\end{figure}

In Fig.~\ref{curL4leg}, we show the dependence of various quantities
as a function of the AB flux $\Phi$ on the $16\times 4$ ring, $n=0.5$,
and $V_{SO}/t=0.6$. In Fig.~\ref{curL4leg}(a) the current of spin-up 
electrons is shown for each chain. Following the behavior
above observed for the $W=3$ strip, the currents no longer satisfy the
conditions $J_{\uparrow,y}=-J_{\uparrow,4-y+1}$ as soon as the AB flux
is nonzero,
but the relations $J_{\downarrow,y}=J_{\uparrow,4-y+1}$ still hold. The SO
currents on each chain are shown in Fig.~\ref{curL4leg}(b), where it
can be verified the symmetry condition, $J_{so,y}=J_{so,4-y+1}$.
The total $S^z(y)$ in each chain, which provides a measure of partial
spin accumulation on each chain, is shown in Fig.~\ref{curL4leg}(c).
It should be noticed that the largest difference in $S^z(y)$ occurs
between the symmetrical inner chains  $y=2,3$. This largest contribution
to the total spin accumulation from the inner chains occurs for all
$V_{SO}/t$ (except the pathological case of $V_{SO}/t=0.8$) as it 
will be shown in Fig.~\ref{all4lsoU}(c). Finally, in 
Fig.~\ref{curL4leg}(d) we show the total spin accumulation, defined
as $\Delta S^z=S^z(1)+S^z(2)-(S^z(3)+S^z(4))$, confirming that its
maximum value around $\Phi=0.4$ is mainly originated by the 
difference in $S^z(y)$ in the inner legs.

In Fig.~\ref{all4lsoU}(a), the flux-averaged spin-up electron currents
on the external and internal legs of the four-chain ring are shown as
a function of $V_{SO}/t$ and for the values of $U$ indicated on the
plot. To be more precise, we plot the difference 
$J_{\uparrow,y}-J_{\uparrow,4-y+1}$, or equivalently, 
$J_{\uparrow,y}-J_{\downarrow,y}$, $y=1,2$. This average is
equal to $J_{\uparrow,y}$ at $\Phi=0$. The behavior of the flux 
averaged differences have the same behavior of
$J_{\uparrow,y}(\Phi=0)$, as it can be assessed by comparing it with
Fig.~\ref{figvso23}(c) (differences between the results in both
Figures are just finite size effects). It is
interesting to notice that the current on the inner leg has the
opposite sign of the current on the outer legs, and that both
currents reach their maximum absolute value at $V_{SO}/t\sim0.5$.
After this point the inner current starts to increase becoming positive
at $V_{SO}/t\sim0.7$ while the outer current decreases. At 
$V_{SO}/t\sim0.8$, the current in the inner leg becomes larger than the
one of the outer leg.

In Fig.~\ref{all4lsoU}(b) SO currents in inner and outer legs are 
shown. In spite of an irregular behavior due to the finite size
involved in the calculations, it can be seen observed that SO currents
roughly follow the behavior of the up currents in the inner leg but
their behavior is opposite for the outer legs. More relevant for
spintronics applications is the behavior observed for the absolute
value of $\Delta S^z_n=S^z_n-S^z_{4-n+1}$, shown in 
Fig.~\ref{all4lsoU}(c). It can be observed that $\Delta S^z_2$, 
corresponding to the inner legs, definitely dominates over
$\Delta S^z_2$, corresponding to the outer legs, and determines
the overall behavior of the resulting spin accumulation,
$\Delta S^z$. The absolute value of this quantity is plotted in
Fig.~\ref{all4lsoU}(d), and it can be observed that its maximum is
obtained at intermediate values of $V_{SO}/t$, and also in 
agreement with previous calculations on the 2-leg ladder\cite{riera},
it is enhanced by the Hubbard repulsion U.

\subsection{Voltage bias}
\label{voltdmrg}

In this subsection we present results obtained with DMRG on 
strips (or generalized ''ladders") with two and three coupled 
chains (or ''legs") with OBC in both directions. The purpose
is to determine the time evolution of the Rashba helical currents
following the sudden application of a bias voltage between both
halves of the strip, as explained in Section~\ref{modelmethod}.

\begin{figure}[t]
\includegraphics[width=0.9\columnwidth,angle=0]{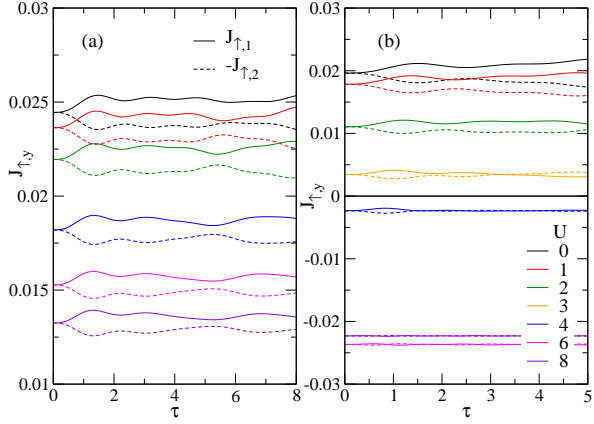}
\caption{[Color Online] Time evolution of the currents of spin-up 
electrons on each chain of a 2-chain strip for various values of $U$
indicated on the plot, (a) $V_{SO}/t=0.6$, $n=0.5$, (b)
$V_{SO}/t=0.4$, $n=1$. DMRG results for the $24\times 2$ strip with
OBC. For the sake of clarity, $J_{\uparrow,2}=J_{\downarrow,1}$ (see
text) has been inverted.}
\label{curlegtim}
\end{figure}
Similarly to the previous case, the applied voltage bias installs a
net hopping current in the longitudinal $x$-direction, $J_{hop}$,
together with a spin-flipping current, $J_{SO}$, and a total
current, $J_{tot}=J_{hop}+J_{SO}$.
This net SO current implies in turn that the two contributions
$J'_{SO,x,y,\hat{x}}$ and $J''_{SO,x,y,\hat{x}}$
will no longer cancel each other, and as a consequence, the
relation on each link,
$J_{\uparrow,x,y,\hat{\mu}}=-J_{\downarrow,x,y,\hat{\mu}}$
no longer holds.
The antisymmetry of the system still implies that 
\begin{eqnarray}
J_{\uparrow,x,y,\hat{x}}=-J_{\downarrow,x,W-y+1,\hat{x}},
\label{asymupdown}
\end{eqnarray}
$y=1,\ldots,W$.
Of course, no net SO current, which is a {\em charge} current,
in the transversal direction will appear.
A more subtle analysis would show the existence of 
transversal {\em spin} currents\cite{sinova04}, leading in
turn to the spin accumulation that we will describe below.

As in Subsection~\ref{noninter.open}, the currents on each chain are
measured at the center of each chain, $x=L/2$, and taking advantage
of the condition Eq.~(\ref{asymupdown}), only the currents
 $J_{\uparrow,l,\hat{x}}$ will be shown in the figures below.

Let us start to examine the case of a strip with $W=2$, or "two-leg
ladder". Although some results for the total hopping or
spin-conserving, and SO or spin-flipping currents, were already
reported,\cite{riera} the main goal here is to explain their behavior
in terms of the RHC studied in the previous Section, which are only 
slightly modified after applying the external voltage bias.

Fig.~\ref{curlegtim} shows the time-evolution of the up-spin electron
currents on both chains of a $24\times 2$ strip, following the
application of a voltage bias at time $\tau=0$. 
Figs.~\ref{curlegtim}(a) and (b) correspond to electron fillings
$n=0,5$ and $n=1$ respectively. For the
sake of clarity, we have inverted one of them, $J_{\uparrow,2}$, since 
$J_{\uparrow,2}=-J_{\uparrow,1}=J_{\downarrow,1}=-J_{\downarrow,2}$
in the static case, that is, at the start of the time evolution.
As it can be observed, as $\tau$ increases, both currents evolve
following different oscillatory behaviors around their $\tau=0$
values. This difference
leads to a nonzero total hopping current along each chain, 
$J_{\uparrow,1}+J_{\downarrow,1}$, and a net total current on the
strip, as it will be further discussed below. During the
whole time evolution the spin-down electron currents on each leg 
remain exactly equal to their spin-up counterpart on the
{\em opposite} leg, that is $J_{\uparrow,2}=J_{\downarrow,1}$, 
$J_{\uparrow,1}=J_{\downarrow,2}$.
Hence, on each chain there will appear a nonzero polarized
current, $J_{pol,l}=J_{\uparrow,l}-J_{\downarrow,l}$, $l=1,2$
that cancel each other leading to an overall zero polarized
hopping current along the strip.

\begin{figure}[t]
\includegraphics[width=0.9\columnwidth,angle=0]{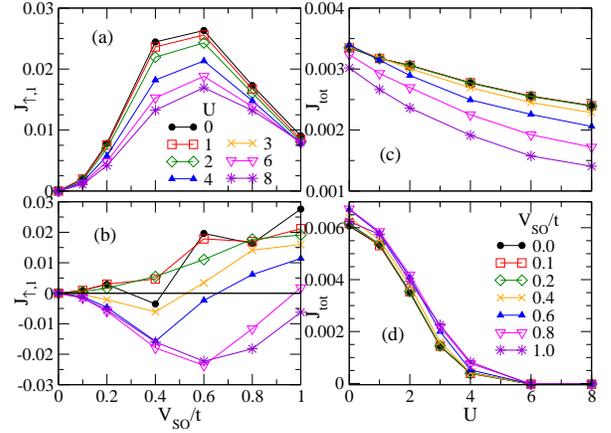}
\caption{[Color Online] Current of spin-up electrons as a function of
$V_{SO}/t$ and various values of $U$ indicated on the plot, for (a)
$n=0.5$ and (b) $n=1$. Total current as a function of $U$ and
various values of $V_{SO}/t$ indicated on the plot,  for (c)
$n=0.5$ and (d) $n=1$. Results obtained with DMRG on the 
$24\times 2$ strip with OBC.}
\label{cur2ldmrg}
\end{figure}

For both densities considered, the effect of the Hubbard correlation is
to suppress both the RHC currents as well as their difference. For
$n=0.5$ (Fig.~\ref{curlegtim}(a)) the system remains metallic for all
$U$ considered, as it can be inferred from (Fig.~\ref{cur2ldmrg}(c)),
and the RHC are smoothly decreasing while keeping their sign. 
On the other hand, at half-filling, (Fig.~\ref{curlegtim}(b)) the RHC,
as well as their differences, decrease more rapidly, and at 
$U \approx 3$ the RHC change their sign and they virtually cancel each
other, leading to a vanishing overall current (Fig.~\ref{cur2ldmrg}(d)).

The result of a systematic analysis of the behavior illustrated in 
Fig.~\ref{curlegtim} for the  $24\times 2$ strip is presented  in 
Fig.~\ref{cur2ldmrg}. At $n=0.5$, $U=0$, the current of spin-up 
electrons in leg "1" at $\tau=0$ (Fig.~\ref{cur2ldmrg}(a)), which
as discussed in the previous paragraph is approximately equal
to $J_{pol,1}/2$, follows
the general behavior illustrated in Fig.~\ref{figvso23}(a) for rings, 
with an approximate initial quadratic behavior as a function of
$V_{SO}/t$, and becoming increasingly suppressed for $V_{SO}/t> 0.6$.
The effect of the Hubbard repulsion is to suppress these RHC, and 
this suppression is more pronounced near $V_{SO}/t=0.5$. The total
currents for this filling, shown in Fig.~\ref{cur2ldmrg}(c) for 
various values of $V_{SO}/t$, are suppressed following the relation
predicted for Luttinger liquids,\cite{kanefisher}
$J_{tot}(U)/J_{tot}(0)=K_\rho$ where $K_{\rho}$ is the one computed
for Hubbard two-leg ladders\cite{orignac} although the Rashba term
could lead to corrections as observed in onedimensional systems.
\cite{rashbaTL,pletyukhov,gritsev}
It is also worth to emphasize the striking similarity between the
behavior of the RHC (Fig.~\ref{cur2ldmrg}(a)), as a function of
$V_{SO}/t$, with the one of the spin accumulation as it can be
seen in Fig.~4(c) of Ref.~\onlinecite{riera}. Of course, for
$U=0$, this
similarity could be eventually traced back to their common
origin in the spin-Hall effect.\cite{sinova04} The fact that their
behavior with increasing $U$ is the opposite is  typical of 
Hubbard models, where magnetic orderings are in general opposed
to transport.

Results at half-filling are strikingly different. The Rashba helical
currents at $\tau=0$, shown in Fig.~\ref{cur2ldmrg}(b), are not only
suppressed by $U$ but they also change their sign for large Hubbard
repulsion. This behavior is somewhat erratic,
particularly for $V_{SO}/t= 0.4$. Notice that for large $U$, RHC
become quite large in absolute value and their behavior is similar
to the one for $n=0,5$.  This behavior of the helical current is
loosely correlated with the vanishing of $J_{tot}$ with $U$
(Fig.~\ref{cur2ldmrg}(d)), which indicates the Mott-Hubbard type
of metal-insulator transition that is present in Hubbard two-leg
ladders close to half-filling.\cite{noack}

\begin{figure}[t]
\includegraphics[width=0.9\columnwidth,angle=0]{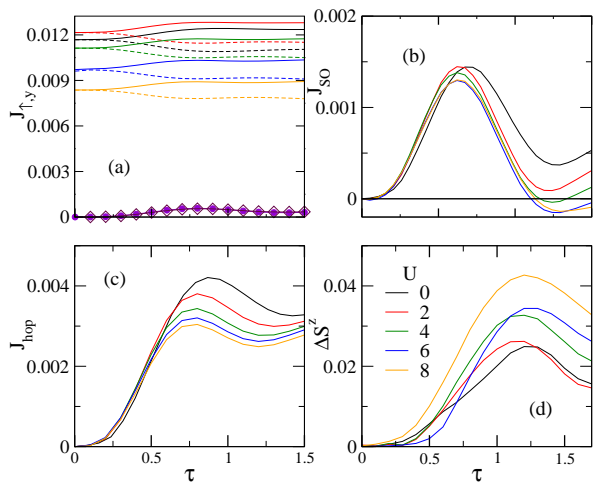}
\caption{[Color Online] Time evolution of (a) $J_{\uparrow,1}$ (full
lines) and $(-J_{\uparrow,3})$ (dashed lines), $J_{\uparrow,2}$,
$J_{\downarrow,2}$ (dotted lines, only for $U=0$), (b) total SO
current, (c) total hopping current (full lines) and total current
(dashed lines), (d) total spin accumulation, for $V_{SO}/t=0.8$,
various values of $U$ indicated on the plot, $n=0.5$. Results
obtained with DMRG on the $16\times 3$ strip with OBC.}
\label{curL3n05so08}
\end{figure}

We try now to see if the previously discussed behaviors could be
also found in wider strips by studying the strip with $W=3$. As it
is well-known, numerical difficulties of applying DMRG to wider
strips increase exponentially, hence most of the results shown
below corresponds to a small strip, $8 \times 3$, and finite size
effects would be estimated by comparing with results for the
$16 \times 3$ strip, although in most cases we could only make
qualitative statements.

Let start with the analysis of the time evolution of the RHC, computed
on the $16\times 3$ strip, shown in Fig.~\ref{curL3n05so08}(a). In
this case, at $\tau=0$ the RHC in the external chains, $y=1$ and 3,
have the same antisymmetric properties as the two chains in the $W=2$
strip, and these currents vanish on the central chain. As in
the $W=2$ case, the voltage bias leads to oscillations of these
currents around their $\tau=0$ values. Relatively small 
differences appearing between antisymmetric currents lead now
to the appearance of a net charge hopping current as well as of a
net charge SO current. The currents in the central chain,
$J_{\uparrow,2}$ and $J_{\downarrow,2}$ evolve with $\tau$ 
around zero but having the {\em same} sign, that is, with net
charge current and vanishing polarized current. The time
evolution of RHC in the external chains, are similar to the ones
observed for the $W=2$ chain, where not only a much stronger 
suppression with $U$ is observed, but also reversal of 
direction.

As discussed above, differences appearing between $J_{\sigma,y}$
currents, $\sigma=\uparrow, \downarrow$, $y=1,\ldots,3$,
lead to net total SO currents, shown in Fig.~\ref{curL3n05so08}(b),
net total hopping currents, shown in Fig.~\ref{curL3n05so08}(c) 
which summed give the total net current. Notice that $J_{hop}$
is, as before, about an order of magnitude smaller
than the RHC, but still it is possible to reliably assign a
plateau value to each of them although it is clearly seen that
they are suppressed by the electron Hubbard $U$ term. The total
SO currents are in turn an order of magnitude smaller than
$J_{hop}$, and clearly, from Fig.~\ref{curL3n05so08}(b), much
more precision and consequent computer power would be required
to assign to them an overall value. However, it seems that it can be
observed a trend with $U$ of not only suppressing these SO currents
but also to reversing its direction, which starts to occur just at  
the end of the time interval considered. This behavior is more clear 
for the smaller $8\times 3$ strip, where larger times can be achieved 
with our currrent computer facilites, and this direction reversal is
consistent, although not as definite, with the behavior observed for 
the $W=2$ strip.\cite{riera}
In the same fashion, it is not reliable to assign a single
value to the evolution of the spin accumulation, which is defined as
$\Delta S^z=S^z_1-S^z_3$, where $S^z_y$ is the sum of all the average
$z$-projection of the spin on each site on chain $y$. However, as it
can be seen in Fig.~\ref{curL3n05so08}(d), results point to a 
trend of enhancement of $\Delta$ with $U$, which again is consistent
with the behavior observed for the $W=2$ strip. However, for 
$V_{SO}/t=0.4$ and 0.6, the behavior is not monotonic with $U$, and
this is another indication that the $W=3$ strip may be not in the
same class of the $W=2$ strip, which is not unexpected. Clearly, far
more computational effort on the $W=3$ strip would be needed to set
this question. Another interesting feature that can be observed in
Fig.~\ref{curL3n05so08} is that $\Delta S^z$ reaches its maximum
with a delay with respect to the time at which $J_{hop}$ and 
$J_{SO}$ reach their maximum. This delay is an indication of the
presence of transversal spin currents appearing after the charge
currents have been installed and leading to the spin accumulation,
as expected in the SHE mechanism with open boundaries.

\begin{figure}[t]
\includegraphics[width=0.9\columnwidth,angle=0]{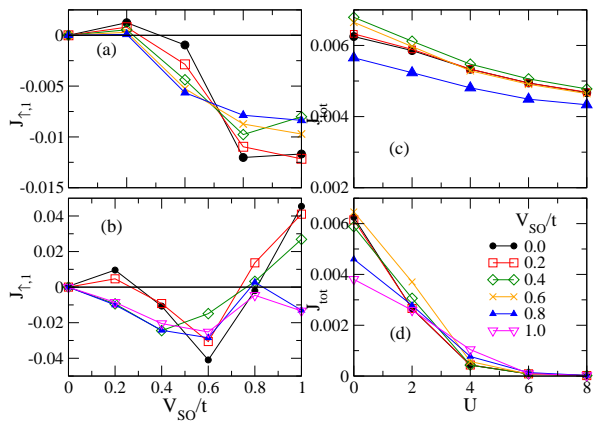}
\caption{[Color Online] Current of spin-up electrons as a function of
$V_{SO}/t$ and various values of $U$ indicated on the plot, for (a)
$n=0.5$ and (b) $n=1$. Total current as a function of $U$ and
various values of $V_{SO}/t$ indicated on the plot,  for (c)
$n=0.5$ and (d) $n=1$. Results obtained with DMRG on the 
$16\times 3$ strip ((a),(c)) and $8\times 3$ ((b),(d)) with OBC.}
\label{cur3ldmrg}
\end{figure}

Finally, a more systematic dependence of the RHC as a function of
$V_{SO}/t$ and $U$ is presented in Figs.~\ref{cur3ldmrg}(a) and  (b),
for $n=0.5$ and  $n=1$, respectively. Although $J_{\uparrow,1}$ have
nonmonotonic variations with $V_{SO}/t$, their variation with
$U$ is noticeable weaker and smoother for  $n=0.5$ than for $n=1$.
Actually, for $n=1$, for most values of $V_{SO}/t$, $J_{\uparrow,1}$
changes its sign as $U$ is increased, similarly to the previous
analyzed case of $W=2$ strips. Again, this behavior occurs while 
the total current, depicted in Fig.~\ref{cur3ldmrg}(d), goes to
zero, indicating a Mott-Hubbard metal-insulation transition.
On the other hand, the behavior of $J_{tot}$, for a given 
$V_{SO}/t$ decreases with $U$ following the general metallic behavior
of quasi-onedimensional systems. We are not aware of Luttinger
Liquid studies on the Hubbard model on $W=3$ strips, and 
consequent estimations of $K_{\rho}$.

\section{Conclusions}
\label{conclusection}

In this manuscript, we have described in the first place the spatial
texture of Rashba helical currents in planar strips, as a function of
various parameters, such as RSOC strength, strip width, electron
filling, in the absence of external electromagnetic sources, and for
the noninteracting case. By numerically solving the single particle
Hamiltonian Eq.~(\ref{Rham}), we found that these RHC are mostly
concentrated at the strip edges, but as the RSOC strength and/or the
width are small enough, they appear as spread over the whole system.
This is a different behavior to the one theoretically predicted for
the helical edge currents, characteristic of the QSH, and topological
insulators, as resulting for example from the Kane-Mele model.
Certainly, this difference could be traced to the difference
between the Haldane-like term for counter propagating spin up and
down electrons, present in that model, which leads
to the opening a gap in the energy spectrum, and hence to an 
insulator state in the bulk, and the Rashba SOC which favours
the metallic state.

The results also suggest a periodicity of the sign oscillations of
the RHC that could originate as Friedel oscillations induced by the
edges, thus explaining its filling dependence. As mentioned in the
Introduction, the natural source of these oscillations are local
impurities, and in fact, the current inhomogeneities reported in
Ref.~\onlinecite{bovenzi} could be of the same nature as the present 
RHC. It would be interesting to perform similar calculations on
strips of the Kane-Mele model, for the sake of comparison.

In the second part of the present work, we consider the effects of
charge currents induced by external electromagnetic sources, which
could be relevant for actual spintronic devices. Our main results are
that the RHC remain in the presence of such induced net currents,
and their differences on each chain with respect to their 
stationary equilibrium values, help to explain how charge currents,
both of the spin conserving or the spin flipping types, are
distributed on each chain of the strip. These RHC also allow us
to predict that on each chain there could be polarized charge
currents, although the total current on the strip would be
unpolarized.\cite{mhliu} Since, as predicted by our handwaving 
model presented in Section~\ref{handwaving} in systems where
Rashba and Dresselhaus SOC have equal strength, the RHC would not
exist, and taking into account the relation between RHC
and the net currents, it would be tempting to propose an
explanation for the vanishing of optical conductivity at the
persistent helix point.\cite{RDcond}

In this study, we also include the
presence of on-site electron correlations given
by the Hubbard repulsion.  In the context of
topological insulators, it has been emphasized that the robustness
of helical edge states with respect to interactions is a result of
time-reversal symmetry, and in particular, the effect of a Hubbard
term added to the Kane-Mele Hamiltonian has been
thoroughly analyzed.\cite{laubach,hohenadler} In the model here
considered, the RHC remain for finite values of $U$, although in
the quasi-onedimensional metallic state, their are suppressed 
roughly following the prediction of Luttinger liquid theory.
In fact, even if the system undergoes a Mott-Hubbard type of 
metal-insulator transition, as detected by the vanishing of the
total current, the RHC are still present.

In principle, the Rashba helical currents would be as difficult to
detect as the helical edge currents of the QSH. For the case of
QSH, it has been shown that experimental evidence for helical
edge currents can be achieved by multiterminal 
probes.\cite{buttiker,roth2009,quay}
However, this scheme depends heavily on the strict edge character
of QSH helical currents, and/or on the quantized nature of these
currents. Clearly, a lot of ingenuity will be needed to adapt
those experimental setups to measure the Rashba helical currents we
observe in the present effort, and particularly the presence of
polarized currents varying across the strip section. In addition, the
study of a dissipationless character of the RHC is out of the scope of
the present work. Notice that our model is dissipationless in the
sense that it is not coupled to external degrees of freedom such
as phonons.

Finally, we would like to emphasize apparent differences with some
previous theoretical studies on this model. The band we consider
is a cosine one, and we considered fillings $n \geq 0.5$, hence it
is not simply reducible to a parabolic band, and, more importantly,
the finite width of our system, and eventually also its finite
length, sets a departure from previous studies on the infinite 
two-dimensional system. Certainly, finite size systems are per se
relevant for nanoscale applications.

\begin{acknowledgments}
The authors are partially supported by the Consejo Nacional de 
Investigaciones Cient\'ificas y T\'ecnicas (CONICET) of Argentina.
Useful discussions with A. Dobry, A. Greco,
L. Lara, and A. Lobos, are gratefully acknowledged.
The authors CJG and JAR acknowledge support from CONICET-PIP No.
11220120100389CO. IJH acknowledges support form CONICET-PIP No. 1060 
and PICT-2014-3290.
\end{acknowledgments}


\begin{thebibliography}{}

\bibitem{wolf} S. A. Wolf, D.D. Awschalom, R.A. Buhrman, J.M. Daughton,
    S. von Molnar, M.L. Roukes, A.Y. Chtchelkanova, D.M. Treger, Science
    \textbf{294}, 1488 (2001).

\bibitem{prinz} G.A. Prinz, Science  \textbf{282}, 1660 (1998).

\bibitem{zutic} I. Zutic, J. Fabian, and  S. Das Sarma, Rev. Mod. Phys.
       \textbf{76}, 323, (2004).

\bibitem{awschalom} D. Awschalom, Physics \textbf{2}, 50
         (2009).

\bibitem{sinovaRMP} J. Sinova, S. O. Valenzuela, J. Wunderlich, C. H.
        Back, and T. Jungwirth, Rev. Mod. Phys. \textbf{87}, 1213

\bibitem{hoffmann13} A. Hoffmann, IEEE Transactions on Magnetics 
         \textbf{49}, 5172 (2013).

\bibitem{rashba} E. I. Rashba, Sov. Phys. Solid State
        \textbf{2}, 1109 (1960); Y. A. Bychkov and  E. I. Rashba,
          JETP Lett. \textbf{39}, 78 (1984).

\bibitem{winkler} R. Winkler, \textit{Spin-Orbit Coupling Effects in
       Two-Dimensional Electron and Hole Systems}
       (Springer, New York, 2003).

\bibitem{dyakonov} M. I. Dyakonov and V. I. Perel, Sov. Phys. Solid 
         State \textbf{13}, 3023 (1972).

\bibitem{Hirsch99} J. E. Hirsch, Phys. Rev. Lett. \textbf{83}, 1834
        (1999).

\bibitem{sinova04} J. Sinova, D. Culcer, Q. Niu, N. A. Sinitsyn, T.
        Jungwirth, and A. H. MacDonald, Phys. Rev. Lett. \textbf{92},
        126603 (2004).

\bibitem{murakami06} S. Murakami, Phys. Rev. Lett. \textbf{97}, 236805
        (2006).

\bibitem{vignale10} G. Vignale, J. Supercond. Nov. Magn. \textbf{23}, 3
        (2010).

\bibitem{nikolic} B. K. Nikoli\'c, S. Souma, L. P. Z\^arbo, and J.
        Sinova, Phys. Rev.  Lett. \textbf{95}, 046601 (2005).

\bibitem{malshukov} A. G. Mal’shukov, L. Y. Wang, C. S. Chu, and K. A.
        Chao, Phys. Rev. Lett. \textbf{95}, 146601 (2005).

\bibitem{bercioux} D. Bercioux and P. Lucignano, Rep. Prog. Phys. 
        \textbf{78}, 106001 (2015).

\bibitem{erlingsson} S. I. Erlingsson, J. C. Egues, and D. Loss,
         Phys. Rev. B \textbf{82}, 155456 (2010).

\bibitem{wenk} P. Wenk and S. Kettemann, Phys. Rev. B \textbf{83},
        115301 (2011).

\bibitem{riera} J. A. Riera, Phys. Rev. B \textbf{88}, 045102 (2013).

\bibitem{gothassaad} F. Goth and F. F. Assaad,  Phys. Rev. B \textbf{90},
        195103 (2014).

\bibitem{hwang} H. Y. Hwang, Y. Iwasa, M. Kawasaki, B. Keimer, N.
       Nagaosa, and Y. Tokura, Nature Mater. \textbf{11}, 103 (2012).

\bibitem{caviglia} A. D. Caviglia, M. Gabay, S. Gariglio, N. Reyren,
       C. Cancellieri, and J.-M. Triscone , Phys. Rev. Lett.
       \textbf{104}, 126803 (2010).

\bibitem{benshalom} M. Ben Shalom, M. Sachs, D. Rakhmilevitch, A.
   Palevski, and Y. Dagan, Phys. Rev. Lett. \textbf{104}, 126802 (2010).

\bibitem{ykim}Y. Kim, R. M. Lutchyn, and C. Nayak, Phys. Rev.
        B \textbf{87}, 245121 (2013).

\bibitem{caprara} S. Caprara, F. Peronaci, and M. Grilli,
          Phys. Rev. Lett. \textbf{109}, 196401 (2012).

\bibitem{liu2014} C. Liu, S.-Y. Xu, N. Alidoust, T.-R. Chang, H. Lin,
        C. Dhital, S. Khadka, M. Neupane, I. Belopolski, G. Landolt,
        H.-T. Jeng, R. S. Markiewicz, J. H. Dil, A. Bansil, S. D. 
        Wilson, and M. Z. Hasan, Phys. Rev. B \textbf{90}, 045127 (2014).

\bibitem{banerjee} S. Banerjee, O. Erten and M. Randeria,
       Nature Phys. \textbf{9}, 626 (2013).

\bibitem{sahin} C. Sahin, G. Vignale, and M. E. Flatt\'e,
        Phys. Rev. B \textbf{89}, 155402 (2014).

\bibitem{khalsa} G. Khalsa, B. Lee, and A. H. MacDonald, Phys. Rev.
        B \textbf{88} 041302, (2013).

\bibitem{bucheli} D. Bucheli, M. Grilli, F. Peronaci, G. Seibold, and
        S. Caprara, Phys. Rev. B \textbf{89} 195448 (2014).

\bibitem{seiji15} Z. Zhong, L. Si, Q. Zhang, W-G Yin, S. Yunoki and K.
        Held, Adv. Mater. Interfaces, \textbf{2}, 5 (2015).

\bibitem{murakami} S. Murakami, N. Nagaosa, and S.-C. Zhang, Science
        \textbf{301}, 1348 (2003).

\bibitem{bernevig}  B. A. Bernevig and S. C. Zhang, Phys. Rev. Lett.
        \textbf{96}, 106802 (2006).

\bibitem{sheng} D. N. Sheng, Z. Y. Weng, L. Sheng, and F. D. M.
        Haldane, Phys.  Rev. Lett. \textbf{97}, 036808 (2006).

\bibitem{konig} M. K\"onig, S. Wiedmann, C. Brüne, A. Roth, H.
        Buhmann, L. W. Molenkamp, X.-L. Qi, S.-C. Zhang,
        Science \textbf{318}, 766 (2007).

\bibitem{kanemele} C. L. Kane and E. J. Mele, Phys. Rev. Lett.
        \textbf{95}, 146802 (2005).

\bibitem{okudakimura} T. Okuda and A. Kimura, J. Phys. Soc. Jpn. 
        \textbf{82}, 021002 (2013).

\bibitem{nomura} K. Nomura, J. Sinova, N. A. Sinitsyn, and A. H. MacDonald,
        Phys. Rev. B \textbf{72}, 165316 (2005).

\bibitem{nomura2} K. Nomura, J. Wunderlich, Jairo Sinova, B. Kaestner,
         A. H. MacDonald, and T. Jungwirth
        Phys. Rev. B \textbf{72}, 245330 (2005).

\bibitem{giantSO} M. Sakano, J. Miyawaki, A. Chainani, Y. Takata, T.
        Sonobe, T. Shimojima, M. Oura, S. Shin, M. S. Bahramy, R. Arita,
         N. Nagaosa, H. Murakawa, Y. Kaneko, Y. Tokura, K. Ishizaka,
        Phys. Rev. B \textbf{86}, 085204 (2012).

\bibitem{pareek} T. P. Pareek and P. Bruno, Phys. Rev. B \textbf{65},
        241305 (2002).

\bibitem{sinova14} J. Zelezny, H. Gao, K. Vyborny, J. Zemen, J. Masek,
      A. Manchon, J. Wunderlich, J. Sinova, and T. Jungwirth, Phys. Rev.
      Lett. \textbf{113}, 157201 (2014).

\bibitem{kato} Y. Kato, R. C. Myers, A. C. Gossard, and D. D.Awschalom,
      Science \textbf{306}, 1910 (2004).

\bibitem{ganichev} S. D. Ganichev and L. E. and Golub, Phys. Status
      Solidi B \textbf{251} 1801 (2014).    

\bibitem{splettstoesser} J. Splettstoesser, M. Governale, and U. Zulicke,
         Phys. Rev. B {\bf 68}, 165341 (2003).
      
\bibitem{michetti} P. Michetti and P. Recher, Phys. Rev. B \textbf{83},
       125420 (2011).      

\bibitem{honghirsch} X. Q. Hong and J. E. Hirsch, Phys. Rev. B {\bf 41}, 
        4410 (1990).

\bibitem{giamarchi} T. Giamarchi and C. Lhuillier, Phys. Rev. B {\bf 42}, 
        10641 (1990).

\bibitem{schollwock} U. Schollw\"ock, Rev. Mod. Phys. \textbf{77}, 259
        (2005).

\bibitem{feiginwhite} A. E. Feiguin and S.R.White, Phys. Rev. B
       \textbf{72}, 020404 (2005).

\bibitem{alhassanieh}  K. A. Al-Hassanieh, A. E. Feiguin, J. A. Riera,
       C. A. Busser, and E. Dagotto, Phys. Rev. B {\bf 73}, 195304
       (2006).

\bibitem{MWNM-09} S. R. Manmana, S. Wessel, R. M. Noack, and A. Muramatsu,
        Phys. Rev. B {\bf 79}, 155104 (2009).

\bibitem{schmitteckert} P. Schmitteckert, Phys. Rev. B \textbf{70},
      121302(R), (2004).

\bibitem{bernevig2} B. A. Bernevig, J. Orenstein, and S. C. Zhang, Phys.
        Rev. Lett. \textbf{97}, 236601 (2006).

\bibitem{dresselhaus} G. Dresselhaus, Phys. Rev. \textbf{100}, 580
        (1955).

\bibitem{bovenzi} N. Bovenzi, F. Finocchiaro, N. Scopigno, D. Bucheli,
       S. Caprara, G. Seibold and M. Grilli, J.
       Supercond. Nov. Magn. \textbf{28}, 1273 (2015).

\bibitem{mzhasan14} M. Z. Hasan, S.-Y. Xu and M. Neupane,
         arXiv:1406.1040.

\bibitem{sunzhu} Q. Sun, G.-B. Zhu, W.-M. Liu, A.-C. Ji, Phys. Rev. A
         \textbf{88}, 063637 (2013).

\bibitem{baum2015} Y. Baum, T. Posske, I. C. Fulga, B. Trauzettel,
        and A. Stern, Phys. Rev. Lett. \textbf{114}, 136801 (2015).

\bibitem{dzero} M. Dzero, K. Sun, V. Galitski, and P. Coleman, Phys. Rev.
          Lett. \textbf{104}, 106408 (2010).

\bibitem{ynagaosa} B.-J. Yang and N. Nagaosa,
         Nature Communications \textbf{5}, 4898 (2014).

\bibitem{meza} G. A. Meza and J. A. Riera, Phys. Rev. B \textbf{90}, 
       085107 (2014).

\bibitem{laubach} M. Laubach, J. Reuther, R. Thomale, and S. Rachel,
         Phys. Rev. B {\bf 90}, 165136 (2014).

\bibitem{mhliu} M. H. Liu, S.-H. Chen, and C.-R. Chang,
         Phys. Rev. B \textbf{78}, 165316 (2008).

\bibitem{wong10} A. Wong and F. Mireles,  Phys. Rev. B \textbf{81}.
         085304 (2010).

\bibitem{kanefisher} C. L. Kane and M. P. A. Fisher, Phys. Rev. B
        \textbf{46}, 7268 (1992).

\bibitem{orignac} E. Orignac and T. Giamarchi, Phys. Rev. B \textbf{56},
         7167 (1997).

\bibitem{rashbaTL} W. Hausler, Phys. Rev. B \textbf{63}, R121310
        (2001).

\bibitem{pletyukhov} M. Pletyukhov and V. Gritsev,
        Phys. Rev. B \textbf{70}, 165316 (2004).

\bibitem{gritsev} V. Gritsev, G. I. Japaridze, M. Pletyukhov, and D. 
        Baeriswyl, Phys. Rev. Lett. \textbf{94}, 137207 (2005).

\bibitem{noack} R. M. Noack, S. R. White, and D. J. Scalapino,
      Physica C \textbf{270}, 281 (1996).

\bibitem{RDcond} Z. Li, F. Marsiglio, and J. P. Carbotte,
         Sci. Rep. \textbf{3}, 2828 (2013).

\bibitem{hohenadler} M. Hohenadler and F. F. Assaad,
         Phys. Rev. B \textbf{90}, 245148 (2014).

\bibitem{buttiker} M. Buttiker, Science \textbf{325}, 278 (2009).

\bibitem{roth2009} A. Roth, C. Brüne, H. Buhmann, L. W. Molenkamp, J,
       Maciejko, X.-L. Qi, S.-C. Zhang, Science \textbf{325}, 294 (2009).

\bibitem{quay} C. H. L. Quay, T. L. Hughes, J. A. Sulpizio, L. N. Pfeiffer,
      K. W. Baldwin, K. W. West, D. Goldhaber-Gordon and R. de Picciotto,
      Nature Physics \textbf{6}, 336 (2010).

\end{thebibliography}
\end{document}